# Silicone-made Tactile Actuator Integrated with Hot Thermo-fiber Finger Sleeve

Mohammad Shadman Hashem [1], Ahsan Raza [2], and Seokhee Jeon [3]

1,2, 3 Department of Computer Engineering, Kyung Hee University, Yongin-si, South Korea

(Email: jeon@khu.ac.kr)

**Abstract ---** Multi-mode haptic feedback is essential to achieve high realism and immersion in virtual environments. This paper proposed a novel silicone fingertip actuator integrated with a hot thermal fabric finger sleeve to render pressure, vibration, and hot thermal feedback simultaneously. The actuator is pneumatically actuated to render a realistic and effective tactile experience in accordance with hot thermal sensation. The silicone actuator, with two air chambers controlled by pneumatic valves connected to compressed air tanks. Simultaneously, a PWM signal from a microcontroller regulates the temperature of the thermal fabric sleeve, enhancing overall system functionality. The lower chamber of the silicone actuator is responsible for pressure feedback, whereas the upper chamber is devoted to vibrotactile feedback. The conductive yarn or thread was utilized to spread the thermal feedback actuation points on the thermal fabric's surface. To demonstrate the actuator's capability, a VR environment consisting of a bowl of liquid and a stove with fire was designed. Based on different functionalities the scenario can simulate the tactile perception of pressure, vibration, and temperature simultaneously or consecutively.

## 1 Introduction

Virtual reality (VR) replicates the real-life experience through visual and auditory stimuli [1]. However, it is challenging to mimic the real-life experience through visual and auditory stimuli only in the VR environment. Enhancing realism and immersion requires the incorporation of haptic feedback, which combines visual and auditory stimuli with physical interactions [2]. Haptic feedback renders different types of physical interaction through tactile sensations such as pressure, vibration, temperature, friction, etc., and kinesthetic sensations such as force and torque [3]. Our work mainly focuses on tactile sensations only. To render different tactile sensations of the virtual objects distinctly in high-fidelity VR environment multi-mode tactile actuators are required [4].

In recent studies, researchers developed tactile actuators with only a single sensation or multi-mode sensations by utilizing different actuators together [5], [6], [7]. However, a single modality reduces the fidelity of the VR environment, while multiple actuators add weight and make the system bulky.

To address these issues, we proposed a dual-layer silicone actuator incorporating a lightweight thermal fabric sleeve for fingertip that simultaneously renders heat, pressure, and vibration feedback. The proposed actuator consists of a silicone-made dual air chamber and a piece of lightweight thermal fabric as a finger sleeve as shown in Figure 1.

This actuator offers significant potential in areas such as safety training, medical education, and rehabilitation, enhancing the realism of virtual training environments. For example, in safety training, it could be utilized to instruct individuals on handling hot objects in industrial settings or to teach children about the risks associated with common household items. In the field of medical training, this technology could greatly aid physicians by replicating the diverse thermal attributes, textures, and stiffness of internal human organs, thereby enhancing the effectiveness of surgical simulations.

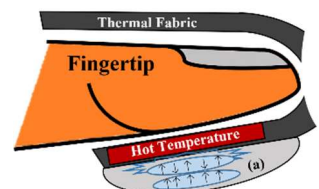

Fig. 1 Proposed fingertip actuator. (a) Silicone layer with dual air chambers

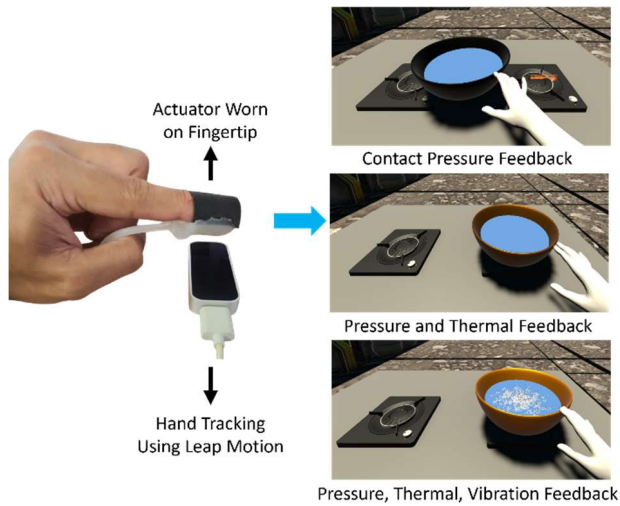

Fig. 2 Illustration of the overall system

## 2 Demonstration

As an example of maximizing the effectiveness of the proposed actuator, we introduce a virtual scenario consisting of a bowl of liquid and a stove with fire to show the perception of contact pressure, and thermal and vibration feedback simultaneously based on different functionalities. The scenario was designed using the Unity Game Engine. In the virtual environment, the hands of the users were tracked using the leap motion sensor.

In the VR scenario, users initially touch a bowl filled with liquid at a normal temperature, experiencing only contact pressure feedback. When the bowl is placed on a stove and the heat is turned on, the bowl changes color and the liquid heat up. At this stage, users experience both contact pressure and thermal feedback simultaneously. As the liquid begins to boil, users perceive simultaneous pressure, thermal, and vibration feedback. The time it takes for the liquid to begin boiling varies depending on the type of liquid. Figure 2 illustrates this entire VR scenario.

## 3 Conclusion

This paper proposed a silicone-based fingertip actuator integrated with a lightweight hot thermal fabric sleeve that renders multiple tactile sensation by providing various types of haptic feedback. The actuator rendered pressure, vibration and thermal feedback simultaneously by utilizing the pneumatic actuation and PWM signal from microcontroller. Finally, to demonstrate the effectiveness of the proposed actuator we designed a virtual scenario with a bowl of liquid and stove with fire. This allows users to experience insignificant temperature variations and tactile sensations comparable to those found in the real world.

### Acknowledgement

This research was supported by the IITP under the Ministry of Science and ICT Korea through the IITP program No. 2022-0-01005 and under the metaverse support program to nurture the best talents (IITP-2024-RS-2024-00425383).